\documentclass{acm_proc_article-sp}

\usepackage{algorithm}                  
\usepackage[noend]{algorithmic}         

\begin{document}

\title{ Compression of Video Tracking and Bandwidth Balancing Routing in Wireless Multimedia Sensor Networks
\titlenote{Permanent address of Yin Wang: Lawrence Technological University, Southfield, Michigan 48075, USA. This work was conducted when Yin Wang was visiting the Southwest Petroleum University.}
}
\numberofauthors{6} 
%
\author{
%
%
\alignauthor
Yin Wang
       \affaddr{School of Computer Science}\\
       \affaddr{Southwest Petroleum University}\\
        \affaddr{Chengdu, Sichuan 610500, China}\\
              \email{ywang12@ltu.edu}
\alignauthor
Jianjun Yang
       \affaddr{Department of Computer Science and Information Systems}\\
       \affaddr{University of North Georgia}\\
       \email{jianjun.yang@ung.edu}
\alignauthor Ju Shen
       \affaddr{Department of Computer Science}\\
       \affaddr{University of Dayton}\\
       \email{ jshen1@udayton.edu}
\and  
\alignauthor Juan Guo
       \affaddr{Department of Computer Science and Information Systems}\\
       \affaddr{University of North Georgia}\\
       \email{juan.guo@ung.edu}
\alignauthor Kun Hua
       \affaddr{Department of Electrical and Computer Engineering}\\
       \affaddr{Lawrence Technological University}\\
       \email{khua@ltu.edu}
}


\maketitle
\begin{abstract}
There has been a tremendous growth in multimedia applications over wireless networks. Wireless Multimedia Sensor Networks(WMSNs)
have become the premier choice in many research communities and industry. Many state-of-art applications, such as surveillance, traffic monitoring, and remote heath care are essentially video tracking and transmission in WMSNs. The transmission speed is constrained by
big size of video data and fixed  bandwidth allocation in constant routing path.
In
this paper, we present a CamShift based algorithm to compress the tracking of videos. Then we propose a bandwidth  balancing strategy in which
each sensor node is able to dynamically select the node for next hop with the highest potential bandwidth
capacity to resume communication. Key to the strategy is
that each node merely maintains two parameters that
contains its historical  bandwidth varying trend and
then predicts its near future bandwidth capacity. Then forwarding
 node  selects the next hop with the highest
potential bandwidth capacity. Simulations demonstrate that our
approach significantly increases the data received by sink node and
decreases the delay  on video transmission in Wireless Multimedia Sensor Network
environment.
\end{abstract}

%


\section{Introduction}
Wireless Multimedia Sensor Networks(WMSNs) have emerged as one of
the key technologies for wireless communications. They
are undergoing rapid development and have inspired numerous
applications because of their advantages. More recently, the availability of inexpensive
hardware such as CMOS cameras and microphones
that are able to ubiquitously capture multimedia
content from the environment has fostered the
development of WMSNs, which are networks of wirelessly
interconnected devices that allow retrieving video
and audio streams, still images, and scalar sensor
data\cite{Ian:survey}.

Wireless multimedia sensor networks will enable several
new applications\cite{Ian:application}, which include:
First, Surveillance.Video and audio sensors will be used
to enhance and complement existing surveillance
systems against crime and terrorist attacks. Large scale networks of video sensors can extend the
ability of law-enforcement agencies to monitor
areas, public events, private properties, and
borders. Second, Traffic Monitoring and Enforcement.It will be possible
to monitor car traffic in big cities or highways and
deploy services that offer traffic routing advice to
avoid congestion.Third, Personal and Health Care. Multimedia sensor
networks can be used to monitor and study the
behavior of  people as a means to identify
the causes of illnesses. Fourth, Environmental and Industrial.Several projects on
habitat monitoring that use acoustic and video
feeds are being envisaged, in which information
has to be conveyed in a time-critical fashion.

Despite of the fact that wireless bandwidth
has been increasing significantly, from a theoretical limit
of 11Mbps for 802.11b to 54 Mbps for 802.11g, and
to 540 Mbps for 802.11n, there has always been high
bandwidth demand resulting from multimedia data\cite{Dong:limit}.
Moreover, the routing path from a node to the sink node is normally created by
a kind of routing protocol and  the path is fixed. Hence some nodes are affording
busy traffic but other nodes that are not in the routing path are idle. For video tracking,
video data is frequently transferred over network. It is extremely important to compress video data and find best transmission strategy.
In this paper, we developed a novel  compression video tracking and bandwidth balancing mechanism to boost the video transmission
 speed in WMSNs. Our contributions are two folds. First, we present a CamShift based algorithm
to compress tracking of videos. Then we propose a bandwidth balancing
strategy in which each sensor node is able to dynamically select the node for next hop with the highest potential
bandwidth capacity to resume communication.
Our approach significantly increase the data received by sink node and decrease the latency on
video transmission in Wireless Multimedia Sensor Network
environment.

The rest of the paper is organized as follows. Section
II discusses the related research on this topic. Section III
proposes a novel method that select the best relay station.
We evaluate the proposed schemes by simulations and
describe the performance results in Section IV. Section V
concludes the paper.

\section{Related Work}
Various approaches regarding video processing \cite{ShenJ:p1,ShenJ:p2,ShenJ:p3,Yang:p1} and transmission \cite{XuK:p1,XuK:p2,XuK:p3,Yang:p4,Yang:p5} over WMSNs were proposed. Q. Cai et al. \cite{Qiao:p1,Qiao:p2} proposed a dynamic action recognition, which is important to capture videos in
WMSNs. In their approach, a Dynamic Structure Preserving Map
is proposed to effectively recognize objects' actions in video
sequences. They modified and improved the adaptive learning procedure
in self-organizing map (SOM) to capture dynamics of best
matching neurons through Markov random walk. The method
can learn implicit spatial-temporal correlations from sequential action feature sets and preserve the intrinsic topologies
characterized by different human motions.
The mechanism is able to learn low-level features in challenging video data. The projection from high dimensional
action features to low dimensional latent neural distribution
significantly reduces the computational cost and data redundancy in the recognition process.

Ahmed et al. \cite{Ahmed:r1} studied improving the quality of MPEG-4 transmission on wireless using Differentiated Services. They investigated QoS provisioning between MPEG-4 video application and
Diffserv networks. To achieve the best possible QoS, all the components involved in the transmission
process must collaborate. For example, the server must use stream properties to describe the QoS
requirement for each stream to the network. They propose a solution by distinguishing the video data
into important one and less important one. Packets marked as less important are dropped in the first case if there is any congestion, so that the receiver
can regenerate the video with the received important information.

Budagavi et al. \cite{Budagavi:r2} improved the performance of video over wireless channels by multiframe video
coding. The multiframe coder uses the redundancy that exists across multiple frames in a typical video
conferencing sequence so that additional compression can be achieved using their MF-BMC (Multi Frame
- Block Motion Compensation) approach. They modeled the error propagation using the Markov chain,
and concluded that use of multiple frames in motion increases the robustness. Their proposed MF-BMC
scheme has been shown to be more robust on wireless networks when compared to the base-level H.263
codec which uses SF-BMC (Single Frame -BMC).

J. Zhang et al.\cite{Zhang:wild,Zhang:simulation} developed a wildlife monitoring system based on wireless image sensor networks.
In their study,  architecture of the wildlife monitoring
system based on the wireless image sensor networks was presented to overcome the shortcomings of the
traditional monitoring methods. Specifically, some key issues including design of wireless image sensor nodes
and software process development have been studied and presented. A self-powered rotatable wireless infrared image
sensor node and an aggregation node designed for large amounts of data were developed. In
addition, their corresponding software was designed. The proposed system is able to monitor wildlife accurately,
automatically, and remotely in all-weather condition, which lays foundations for applications of wireless image
sensor networks in wildlife monitoring.

Q. Sun et al. proposed wireless sensor monitoring systems \cite{SunQ:p1,SunQ:p2,SunQ:p4}. Compared with conventional image and video based monitoring systems, wireless sensor monitoring systems is an alternative technology for pattern recognition in multimedia field due to its low cost, low data throughput, self-management, high robustness under various conditions. Nowadays, such wireless sensor based sensing systems have been widely investigated, especially in human sensing. Specifically, among various wireless sensors, pyroelectric infrared (PIR) sensor and fiber-optic sensor are the most popular ones in human tracking, human identification, and activity recognition fields. Due to the low data throughput property of wireless sensors, the most challenging issues are efficient information acquisition scheme design and light-weight learning algorithm development. A classical work for information acquisition design with a geometric sampling structure and a pseudo-random visibility modulation is created. Such design can effectively reduce the effects of occlusion and energy variation, thus, enhance information acquisition. Another milestone work proposed a space encoding scheme to avoid the ambiguity generated by bipedal movements of humans. This scheme is able to dramatically reduce the false alarm rate in human localization for fiber-optic sensors. With respect to light-weight learning algorithm design, presented a Bayesian networks and region of interests (ROI) based reasoning method. This method can be applied to any sensor modalities.

J. Campbell et al.\cite{Campbell:r3} presented
IRISNET, a sensor network architecture
that enables the creation of a planetary-scale infrastructure of
multimedia sensors that can be shared by a large number of applications.
To ensure the efficient collection of sensor readings, IRISNET
enables the application-specific processing of sensor feeds on
the significant computation resources that are typically attached to
multimedia sensors. IRISNET enables the storage of sensor readings
close to their source by providing a convenient and extensible
distributed XML database infrastructure. Finally, IRISNET provides
a number of multimedia processing primitives that enable the
effective processing of sensor feeds in-network and at-sensor.

S. Shen et al. developed new sensing technologies based on nanostructured devices, which significantly reduced the physical size and weight of sensor nodes and therefore promoted the flexible and cost-effective deployment of wireless sensor networks\cite{Shenshen:p1,Shenshen:p2,Shenshen:p3}.

\section{problem formulation}
To boost the transmission speed of video data over WMSNs, our mechanism
works on two folds. One is the video tracking and compression. The second is
bandwidth balancing routing.
\subsection{Video tracking and compression}
We present CamShift based tracking algorithm based on Histogram Back-projection\cite{{Lee:a1}} and Mean Shift Algorithm\cite{Yang:a1}.
The mean shift algorithm works well on static probability distributions but not on
dynamic ones. Our approach detects the peak in the probability distribution image by
applying mean shift while handling dynamic distributions by readjusting the search
window size for the next frame based on the $0th$ moment of the current frames
distribution, which allows the algorithm to anticipate object movement to quickly
track the object in the next scene. Specifically, the search area can be restricted
around the last known position of the target, resulting in possibly large computational
savings. In a single image, the process is iterated until convergence an upper bound
on the number of iterations is reached. The detection algorithm is applied to successive
frames of a video sequence to track target. This type of scheme introduces a
feed-back loop, in which the result of the detection is used as input to the next
detection process.

\begin{algorithm}[]
\begin{algorithmic}[1]
\caption{{\em Improved CamShift}} \label{algorithm1} \STATE Set the region of interest of the probability distribution image to the
entire image  \STATE Select an initial location of the Mean Shift 2D search window $W$ (scale and
location) in the first frame. The selected location is the target distribution to be
tracked \STATE Calculate a color probability distribution of the region centered at the Mean
Shift search window, and the region should be slightly bigger than the mean
shift search window \STATE Perform mean shift on the area until suitable convergence ($T$=1). Store the 0$th$
moment (distribution area) and centroid location \STATE The search window for the next frame is centered around the centroid and the
size is scaled by a function of the 0$th$ movement; Go to step 3
\end{algorithmic}
\end{algorithm}
In  algorithm 1, we choose 1.0 as the threshold of convergence  to get a
satisfying tracking result. The location of the search window is used to help track
movement.

\subsection{Bandwidth Balancing Routing}
\begin{figure}[!htp]
\begin{center}
\includegraphics[width=7.0cm]{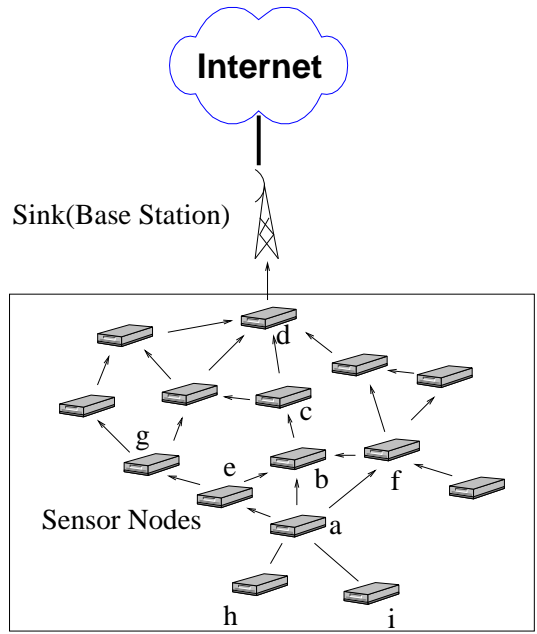}
\end{center}
\caption{An example  of WMSN} \label{wmsn}
\end{figure}

In WMSNs, sensor nodes collect video information, process video, and  transmit  data to sink node(base station).
The sink nodes then sends data to Internet. Fig.~\ref{wmsn} shows an instance of WMSNs, in which sensor nodes handle video data and transfer the data to the sink node, then to Internet. The communications over nodes are based on channels. Channel bandwidth is critical for transmission speed.
We define the residual bandwidth ability(RBA) for a node as its possible bandwidth deducting the real bandwidth over its channel. For Instance, if the possible bandwidth for a node is 100MB/S, the traffic occupies  60MB/S, then its RBA is 40MB/S. We use a simple case to illustrate our main idea. Suppose node $a$ collects video data and got it processed. Then it intends to send the data to the sink node. The routing path can be set up by any routing algorithm. Normally, it is fixed when it is set up. The path a-b-c-d can be a routing path. If many other nodes also send data to $b$, and then to the sink node, $b$ has to afford a lot of traffic, while $e$ or $f$ have low traffic. Considering this scenario, we are thinking about if node $a$ can select one of its neighbors with highest RBA as the next hop.
It may not be $b$.


%
We further assume that the historical RBA vector up to this moment of
 $b$ is [..., 85, 70, 55, 40],    $e$ is [..., 25, 32, 35, 39], and   $f$ is [..., 50, 46, 42, 38].
A simple strategy will let $a$ select $b$ because it  has the highest RBA at the moment.
Seemingly, $b$ is the best choice. However, since $b$ is  a hot node  serving a high volume  traffic and its RBA is going down dramatically, while $e$ or $f$
is serving  less traffic and their RBA are going down slowly or even going up, so $a$  should select $e$ or $f$ to avoid hot node and then balance bandwidth utilization.

Our goal is to let each node  select the node of next hop with highest potential RBA. In our approach,
each candidate node can predict its potential RBA for next time  based on its historical RBA and current RBA. Then the forwarding node selects the best candidate.
In this example, our strategy will let $a$ choose $e$.

\subsection{Calculation of Residual Bandwidth Ability}

The communications between nodes in a Wireless Multimedia Sensor Network vary from time to time. It is critical for a node to forward data to the
 possible best node.
For each node, the trend of its bandwidth's  changing    is essential to calculate  its potential RBA.
Suppose the current time is $k$. A naive method to calculate its potential RBA is this: Let  node $N_{i}$ save
all its historical and current (time $k$) RBA $R_{i_0}$, $R_{i_1}$, ...,$R_{i_k}$  in a vector.
Then it can calculate its potential RBA for time $k+1$ by  curve-fitting in numerical analysis that approximates
its RBA changing trend. The limitations of this method are to spend too much space to save the RBA values in the vector and too complex for curve-fitting computation.

We propose a statistical based strategy to let each node predict  its future RBA. In this scheme, each node only maintains two parameter values. One is its trend of RBA acceleration and the second is the
variance of the acceleration. The two parameters store how its RBA changes. Together with the node's current bandwidth,  its future  bandwidth capacity is achieved by predicted  RBA(potential RBA).
Thus, each node is able to figure out the best next hop subsequently.

We define the following notations to represent the terms regarding our approach for node $N_i$.

\noindent $R_{i_k}$: The measured residual bandwidth at time $k$. \\
 $\hat{R_{i_k}}$: The bandwidth  predicted at time $k$(Potential RBA). \\
$a_{i_k}$: The acceleration measured  at time $k$. It indicates how RBA changes during a time slot.\\
$\hat{a_{i_k}^-}$: The acceleration at time $k$ evolved   from time $k-1$.\\
$\hat{a_{i_k}}$: The potential acceleration at time $k$. \\
 $v_{i_k}$: The variance of acceleration updated  at time $k$.\\
  $v_{i_k}^-$: The variance of acceleration at time $k$   evolved from time $k-1$.\\
$\epsilon$: The error or noise in the process.\\
  $B_{i_k}$: The blending factor at time $k$.\\


\begin{algorithm}[]
\begin{algorithmic}[1]
\caption{{\em Prediction for Future RBA}} \label{algorithm4}
\STATE Measure its current residual bandwidth $R_{i_k}$  \STATE Calculate $a_{i_k}$ as its measured acceleration  by\\
$a_{i_k}$=$(R_{i_k}-R_{i_{k-1}})$/$\Delta$$t$ \STATE Updates $\hat{a_{i_k}^-}$ and $v_{i_k}^-$ in order to keep its historical energy to predict its future energy.\\
$\hat{a_{i_k}^-}=\hat{a_{i_{k-1}}}$ \\
$v_{i_k}^-=v_{{i_k-1}}+\epsilon$
 \STATE Compute the blending factor $B_{i_k}$\\
$B_{i_k}=v_{i_k}^-(v_{i_k}^-+\epsilon)^{-1}=v_{i_k}^-/(v_{i_k}^-+\epsilon)$
  \STATE Calculate its potential acceleration\\
$\hat{a_{i_k}}=\hat{a_{i_k}^-}+B_{i_k}(a_{i_k}-\hat{a_{i_k}^-})$
\STATE Update the variance of acceleration\\
$v_{i_k}=(1-B_{i_k})v_{i_k}^- $
\STATE Predict the RBA of time $k+1$\\
$R_{i_{k+1}}=R_{i_k}+\hat{a_{i_k}}\Delta t $
\end{algorithmic}
\end{algorithm}

At time $k$,  node $N_i$   calls algorithm 2 to calculate future RBA. $N_i$ knows its maintained values $\hat{a_{i_{k-1}}^-}$ and $v_{i_{k-1}}$. It measures its  $R_{i_k}$ in step 1. Then it calculates measured acceleration in step 2.
$N_i$ updates $\hat{a_{i_k}^-}$ and $v_{i_k}^-$ in order to keep its historical RBA to predict its future RBA in step 3.
$N_i$ also computes the blending factor $B_{i_k}$ in step 4,
which indicates how much the acceleration changes from last time to current time.
Once $N_i$ obtains the blending factor $B_{i_k}$ and the evolved acceleration $\hat{a_{i_k}^-}$, it knows how much the acceleration changes with
evolved acceleration. Additionally, $R_i$ considers the measured acceleration $a_{i_k}$. Then it calculates its potential acceleration
in step 5. This acceleration will be used to predict its RBA of time $k+1$. It updates the variance of acceleration for future utilization in step 6 and predicts its future RBA of time $k+1$ in step 7.

\subsection{Region for next hop}

When a node $N_i$ intends to transfer video data to the sink node, it selects the neighbor node with highest potential RBA as its next hop.
Apparently, $N_i$ will not select any nodes in the opposite direction from $N_i$ to the sink node.

\begin{figure}[!htp]
\begin{center}
\includegraphics[width=7.0cm]{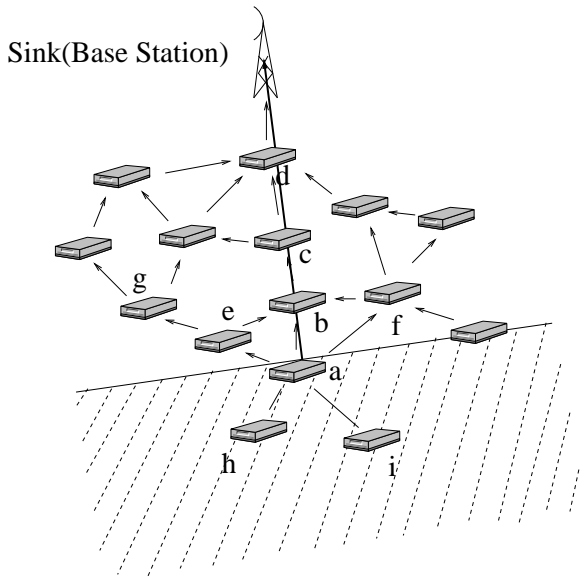}
\end{center}
\caption{Region for next hop} \label{region}
\end{figure}

How does node $N_i$ figure out the region where next hop falls? In current WMSNs, each device is equipped with
GPS and hence it knows  its location. We assume that the sender knows its location and the location of the receiver. The assumption is very common in geographic routing\cite{Yang:p2,Yang:p3}.
Fig.~\ref{region} shows the scenario. Suppose node $a$ intends to send video data to the sink node, it creates a ray by connecting
itself with the sink. Then it draws a line perpendicular to the ray. Then the region on the same side of the sink is available region for next hop. As in Fig.~\ref{region},  $a$ will never select a node in the dash line region for next hop.

\subsection{Bandwidth Balancing Routing}
Once a node $N_i$ intends to send video data to the sink, it needs to choose the node with highest RBA as the next hop. It
 sends a  signal to its neighbors in the available region. Upon receiving the signal, each receiver calls algorithm 2 to calculate its RBA of next time and returns the RBA to  node $N_i$. When  node $N_i$ receives all RBAs, it selects the one with highest RBA as the next hop.

\section{Performance Evaluation}

We evaluated our mechanism in a simulated noiseless
radio network environment by MATLAB. We create a topology that consists of a number of randomly distributed nodes. A sink node is located in the edge of the area. Our approach is based on balancing bandwidth utilization by RBA selection.We compare our approach(``RBA") with LEACH\cite{leach} and GPSR\cite{GPSR}
in terms of data received by sink node and number of alive nodes.
To evaluate the first item, we performed a sequence of experiments in which the number of nodes varies from 50 to 800 with increment of 50. For each number of mobile users, we measure size of data the sink node received 10 times
and present the average.

\begin{figure}[!htp]
\begin{center}
\includegraphics[width=8.0cm]{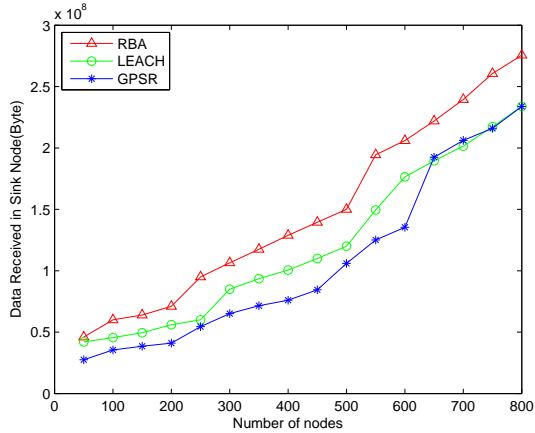}
\end{center}
\caption{Data received by sink node} \label{g22}
\end{figure}

Fig.~\ref{g22} shows our evaluation of data received in sink node. 
Among the three approaches, the scheme GPSR results in the lowest received data. Our approach RBA generate highest received data. This indicates that bandwidth  balancing approach can improve aggregate data  for the sink
because the bandwidths are allocated more reasonable over the network.


\begin{figure}[!htp]
\begin{center}
\includegraphics[width=8.0cm]{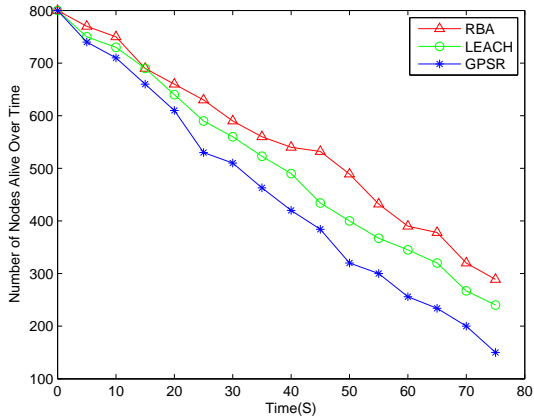}
\end{center}
\caption{Number of alive nodes} \label{g32}
\end{figure}

In WMSNs, the  life time of a relay station is important. If one node runs turns down, there will be a hole\cite{Yang:p2,Yang:p3,GPSR} and the network performance will be
decreased dramatically. So we also measure how many alive nodes along with the time elapsing. We initialize 800 nodes in the environment and observe how many alive nodes every 5 seconds up to 80$th$ second.
Fig.~\ref{g32} shows how many alive nodes of the three schemes with time elapsing.
GPSR results in the lowest average lift time. That is because GPSR adopts   a landmark node to forward a lot of data when there is a hole. However, the landmark is easy to turn down with  high volume traffic. Then the hole is bigger \cite{Yang:p2,Yang:p3}.
Our approach maintains the largest alive nodes  because we always balance bandwidth and then makes the life time longer.

\section{Conclusion}
\label{conclusion}

In this paper, we present a RBA based bandwidth balancing mechanism to transfer video data in
Wireless Multimedia Sensor Networks. We first present a CamShift based approach to compress video for video tracking and transmission.
Then we propose a statistical based mechanism to
balance bandwidth utilization.
In the approach, each node only needs to remain two factors to store its
historical residual bandwidth ability(RBA)  trend and then it is able to predict its bandwidth of near future. Each node selects the
next hop  with
potential highest  RBA to relay data transmission.
Simulations demonstrate that our approach generates highest received data to sink node  in the network and largest alive nodes over time elapsing. 

%
%

%
\bibliographystyle{abbrv}
\bibliography{sigproc}  
\balancecolumns
\end{document}